\documentclass[aps,showpacs]{revtex4}
\usepackage{graphics}
\usepackage{graphicx}

\begin{document}
\title{Rabi oscillations under ultrafast excitation of graphene}
\author{P.N. Romanets}
\author{F.T. Vasko}
\email{ftvasko@yahoo.com}
\affiliation{Institute of Semiconductor Physics, NAS of Ukraine,
Pr. Nauky 41, Kiev, 03028, Ukraine}
\date{\today}

\begin{abstract}
We study coherent nonlinear dynamics of carriers under ultrafast interband
excitation of an intrinsic graphene. The Rabi oscillations of response appear
with increasing of pumping intensity. The photoexcited distribution is calculated
versus time and energy taking into account the effects of energy relaxation
and dephasing. Spectral and temporal dependencies of the response on a probe
radiation (transmission and reflection coefficients) are considered for different
pumping intensities and the Rabi oscillations versus time and intensity are
analyzed.
\end{abstract}


\maketitle
The Rabi oscillations of coherent response under ultrafast excitation of two-level
atomic systems were studied during last decades. \cite{1} Similar phenomena
in bulk semiconductors and quantum wells have also been observed, see \cite{2,3}
and reviews. \cite{4,5} Recently, the properties of graphene under ultrafast
interband excitation were investigated, and most attention has been concentrated
on relaxation dynamics for the case of the linear excitation regime of
epitaxial and exfoliated graphene, see \cite{6} and \cite{7} respectively.
It was found that the relaxation times of photoexcited electron-hole pairs due to
the optical phonon emission are about 0.1 ps \cite{6,7}, so that an
investigation of coherent dynamics is possible during the femtosecond time
scales when the regime of Rabi oscillations can be realized with increasing of
pumping intensity. To the best of our knowledge, neither experimental nor
theoretical treatment of the nonlinear coherent response for the Rabi oscillations
regime in graphene is not performed yet. Due to the gapless and massless energy
spectrum with a neutrinolike dispersion law $\pm v_Wp$ ($v_W\simeq 10^8$ cm/s
is the characteristic velocity for the Weyl-Wallace model \cite{8}), such a case
should be essentially different from the above mentioned cases. \cite{2,3,4,5}

In this paper, we consider the nonlinear coherent process of carrier photoexcitation
in an intrinsic graphene under the slow-envelope condition, $\Omega\tau_p\gg 1$
where $\Omega$ is the frequency of light and $\tau_p$ is the duration of excitation.
Eliminating the nondiagonal components of density matrix, which are responsible for
the high-frequency [$\propto\exp (-i\Omega t)$] oscillations of polarization, we
describe such a process by the same distribution functions for electrons and holes,
$f_{pt}$ because of the electron-hole symmetry. The kinetic equation for $f_{pt}$
takes form
\begin{equation}
\frac{df_{pt}}{dt} =G(f_p|t)+J_0(f_t|p) ,
\end{equation}
where $G(f_p|t)$ is the interband generation rate and $J_0(f_t|p)$ is the collision
integral. The general consideration of Eq. (1) can be found in \cite{9} and the
evaluation of the generation rate for graphene is performed in Ref. 10. Further,
we solve this equation and analyze the spectral and temporal dependencies of the
transmission and reflection coefficients for a probe radiation at different pumping
levels.

The coherent interband photoexcitation caused by the in-plane electric field
$w_t{\bf E}\exp (-i\Omega t)+c.c.$ with the envelope form-factor $w_t$ is described
by the generation rate in Eq. (1) \cite{10}
\begin{eqnarray}
G(f_p|t)=\left( \frac{eEv_W}{\hbar\Omega}\right)^2 w_t
\int\limits_{-\infty}^0 dt' w_{t+t'}e^{t'/\tau_d} \nonumber \\
\times\cos\left[\left(\frac{2v_W p}{\hbar}-\Omega\right)t'\right]
\left( 1-2f_{pt+t'}\right) .
\end{eqnarray}
Here we neglect the multi-photon interband transitions under the condition
$(eEv_W /\hbar\Omega^2)^2\ll 1$ and the dephasing relaxation time $\tau_d$
is introduced phenomenologically. The non-local factor $(1-2f_{pt+t'})$
describes the Pauli blocking and the temporal memory effect which is responsible
for the Rabi oscillations. Since the cascade emission of optical phonons
dominates in the energy relaxation, \cite{6,7,11} Eq. (1) involves the
collision integral
\begin{equation}
J_{0}(f_t|p)=\nu_{p+p_0}(1-f_{pt})f_{p+p_0t}-\nu_{p-p_0}(1-f_{p-p_0t})f_{pt} ,
\end{equation}
where $p_0=\hbar\omega_0/v_W$ is the characteristic momentum, $\hbar\omega_0$
is the optical phonon energy, and $\nu_p=v_0p/\hbar$ is the relaxation rate
for the spontaneous emission of optical phonons. Under the condition $\nu_{p_0}
\tau_p\ll 1$ we neglect interband transitions due to optical phonon emission. We
also restrict ourselves by the model with a single phonon of energy $\hbar
\omega_0\simeq$0.2 eV with the efficiency of coupling determined by the characteristic
velocity $v_0\simeq 10^6$ cm/s. These parameters are in agreement with the
results of calculations of the electron-phonon interaction \cite{12} and with the
measurements of relaxation dynamics. \cite{6,7} The problem given by Eqs. (1)
- (3) is solved below with the initial condition $f_{pt\to -\infty}=0$ which
is correspondent to the undoped graphene.

The analytical solution of the problem formulated can be found for the collisionless
case, $\tau_d\to\infty$ and $v_0\to 0$, under the resonant condition $p\to\hbar\Omega
/2v_W\equiv p_\Omega$, when the oscillating factor is absent in the integral
generation rate (2). For such a case, the integro-differential Eq. (1) can
be transformed into the second order differential equation and the solution takes
form
\begin{equation}
f_{p=p_\Omega ,t} =\frac{1}{2}\left[ 1-\cos\left(\sqrt{2I_{ex}}\int_{-\infty}^t
\frac{dt'}{\tau_p}w_{t'}\right)\right] .
\end{equation}
Here we introduced the dimensionless intensity, $I_{ex}=(eE\tau_p v_W/\hbar\Omega )^2$,
so that the Rabi oscillations of the resonant distribution with time and with field
strength ($\sqrt{2I_{ex}}\propto E$) takes place. At $t\to\infty$ one obtains the
resonant distribution $f_{p=p_\Omega ,t\to\infty}=\left( 1-\cos{\cal A}_{ex}\right) /2$
which is determined by the dimensionless area of the incident pulse ${\cal A}_{ex}
=\sqrt{2I_{ex}}\int_{-\infty}^{\infty}dtw_{t}/\tau_p$.
\begin{figure}[ht]
\begin{center}
\includegraphics{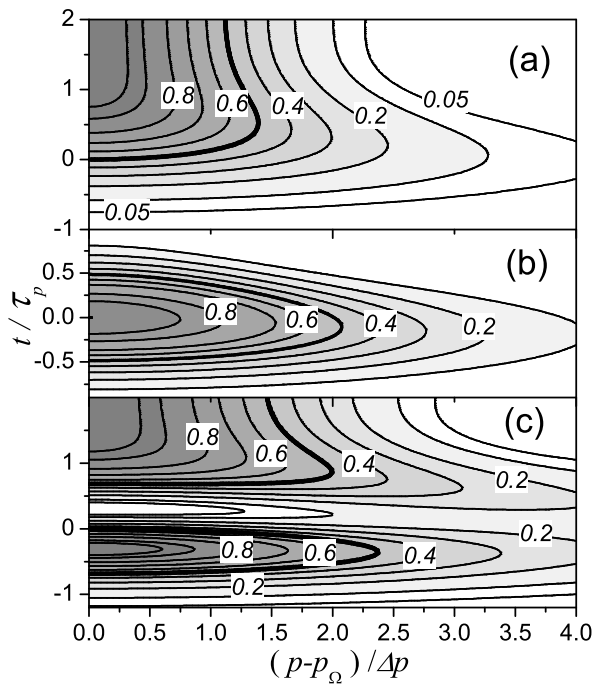}
\end{center}\addvspace{-1 cm}
\caption{Contour plots of photoexcited distribution $f_{pt}$ versus dimensionless
momentum and time, $(p-p_\Omega )/\Delta p$ and $t/\tau_p$ for the collisionless
regime at pumping levels corresponding to ${\cal A}_{ex}=\pi$ (a), $2\pi$ (b) and
$3\pi$ (c). }
\end{figure}

The collisionless case at $p\neq p_\Omega$ is described by the reduced equation (1)
$df_{pt}/dt=G(f_p|t)$. The numerical solution of this equation is obtained here
with the use of the finite difference method \cite{13} and the Gaussian form-factor
$w_t=\sqrt[4]{2/\pi}\exp\left[ -\left( t/\tau_p\right)^2\right]$. \cite{14} In
Fig. 1 we plot $f_{pt}$ versus dimensionless time, $t/\tau_p$, and momentum
$(p-p_\Omega )/\Delta p$ which is centered at $p_\Omega$. Here $\Delta p=\hbar
/(v_W\tau_p )$ determines a width of photoexcited distribution at $t\geq 2\tau_p$
while a width of distribution at $t\simeq 0$ increases with pumping intensity,
as it is shown in Figs. 1 a-c. As $I_{ex}$ increases, a temporal Rabi oscillations
at $p\neq p_\Omega$ are similar to the transient evolution described by Eq. (4).
\begin{figure}[ht]
\begin{center}
\includegraphics{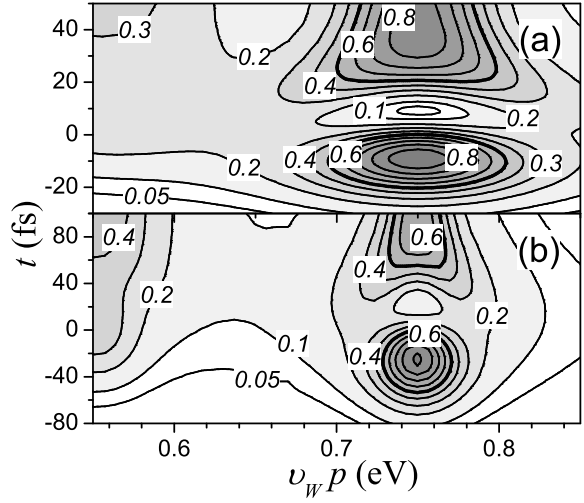}
\end{center}\addvspace{-1 cm}
\caption{Contour plots of photoexcited distributions $f_{pt}$ versus energy $v_Wp$
and time for pulse durations $\tau_p=$30 fs (a) and 80 fs (b) at pumping level
corresponding to ${\cal A}_{ex}=3\pi$.}
\end{figure}

We turn now to consideration of the problem (1) - (3) taking into account the
dephasing and energy relaxation processes, when $\tau_d$ and $v_0$ are
finite. In the calculations below we use the pumping frequency $\hbar\Omega =$1.5 eV
and the dephasing time $\tau_d\simeq\nu_{p_\Omega}^{-1}=$85 fs, which is correspondent
to the process of spontaneous emission of optical phonons. Since the cascade
emission of dispersionless optical phonons, described by Eq. (3),
a multipeak structure of $f_{pt}$ with maxima centered around $p_\Omega -kp_0$
($k=0,1,\ldots$) is realized. The only first phonon repetition is essential during
the photoexcitation process, when $t\leq 2\tau_p$, see Figs. 2a and 2b where the
contour plots of the photoexcited distributions $f_{pt}$ at the pumping level
${\cal A}_{ex}=3\pi$ are shown for the pulse durations $\tau_p=$30 fs and 80 fs.
One can see that the photoexcited distribution appears to be narrower with
increasing of $\tau_p$ and that the Rabi oscillations are visible at $\tau_d
\sim\tau_p$ (Fig. 2b).
\begin{figure}[ht]
\begin{center}
\includegraphics{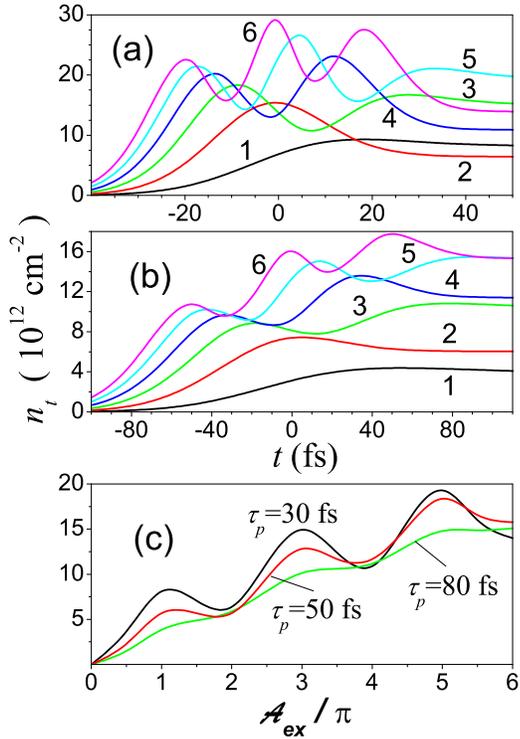}
\end{center}\addvspace{-1 cm}
\caption{(Color online) Transient evolution of concentration of photoexcited carriers
for pulse durations $\tau_p=$30 fs (a) and 80 fs (b) at pumping levels corresponding
to ${\cal A}_{ex}=\pi$ (1), $2\pi$ (2), $3\pi$ (3), $4\pi$ (4), $5\pi$ (5), and
$6\pi$ (6). (c) Photoexcited concentration versus ${\cal A}_{ex}$ at $t\gg\tau_p$
for different $\tau_p$ (marked).}
\end{figure}

Using the distribution $f_{pt}$ obtained, we calculate here the concentration of
photoexcited carriers
\begin{equation}
n_t=\frac{2}{\pi\hbar^2}\int_0^\infty dppf_{pt}
\end{equation}
versus time and versus pumping level at $t\gg\tau_p$.
As one can see from Figs. 3a and 3b, the amplitudes of temporal oscillations decreases
with increasing of $\tau_p$ but the oscillations remain visible at $\tau_p\sim\tau_d$.
The oscillatory behavior of $n_{t\to\infty}$ versus pumping level (which is proportional
to ${\cal A}_{ex}^2$) appears to be suppressed at $\tau_p\sim\tau_d$ as it is shown in
Fig. 3c. The condition ${\cal A}_{ex}=\pi$ corresponds to the pulse energies $\sim$15 nJ
or $\sim$5.6 nJ for $\tau_p=$30 fs or 80 fs and for the sport area $\sim 10^{-4}$ cm$^2$
(note, that the pulse energy $\propto\tau_p^{-1}$ and $\propto n^2$, if ${\cal A}_{ex}
=n\pi$).

\begin{figure}[ht]
\begin{center}
\includegraphics{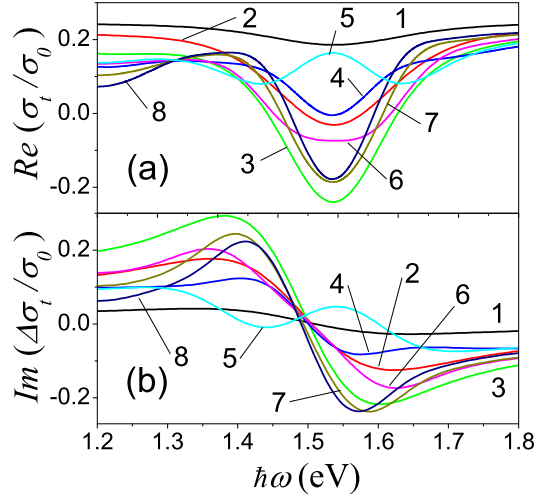}
\end{center}\addvspace{-1 cm}
\caption{(Color online) Spectral dependencies of the dynamic conductivity,
${\rm Re}\sigma_{\omega t}$ (a) and ${\rm Im}\Delta\sigma_{\omega t}$ (b), around the
pumping energy 1.5 eV for delay times between $-$30 and 40 fs with step 10 fs (marked by
1 - 8) at pumping level corresponding to ${\cal A}_{ex}=3\pi$ and pulse duration
$\tau_p =$30 fs.}
\end{figure}

The transient response on a probe radiation $\propto\exp (-i\omega t)$ is determined
by the dynamic conductivity $\sigma_{\omega t}$ which was evaluated in \cite{9,15}
for the collisionless case $\omega\tau_d\gg 1$, when the parametric dependency on
time takes place. The real and imaginary parts of $\sigma _{\omega t}$ are given by
\begin{eqnarray}
{\rm Re}\sigma_{\omega t}=\frac{e^2}{4\hbar}(1-2f_{p_\omega ,t}) , ~~~~ \\
{\rm Im}\sigma_{\omega t}=\overline{\sigma}_{\omega}-\frac{e^2}{\pi\hbar}
{\cal P}\int\limits_0^\infty\frac{dyy^2}{1-y^2}f_{p_\omega y,t} , \nonumber
\end{eqnarray}
where $\cal P$ means the principal value of integral and we introduced the
time-independent contribution $\overline{\sigma}_{\omega}$ described the
undoped graphene in the absence of photoexcitation (below we use the phenomenological
expression for $\overline{\sigma}_{\omega}$ introduced in \cite{16}). The
negative absorption condition ${\rm Re}\sigma_{\omega t}<0$ takes place if
$f_{p_\omega ,t}>1/2$; these regions were marked by thick curves in Figs. 1 and 2.
The peak shape of ${\rm Re}\sigma_{\omega t}$ centered at $\omega =\Omega$
leads to a visible spectral dispersion of ${\rm Im}\sigma_{\omega t}$ at
$\omega <\Omega$ and $\omega >\Omega$.
\begin{figure}[ht]
\begin{center}
\includegraphics{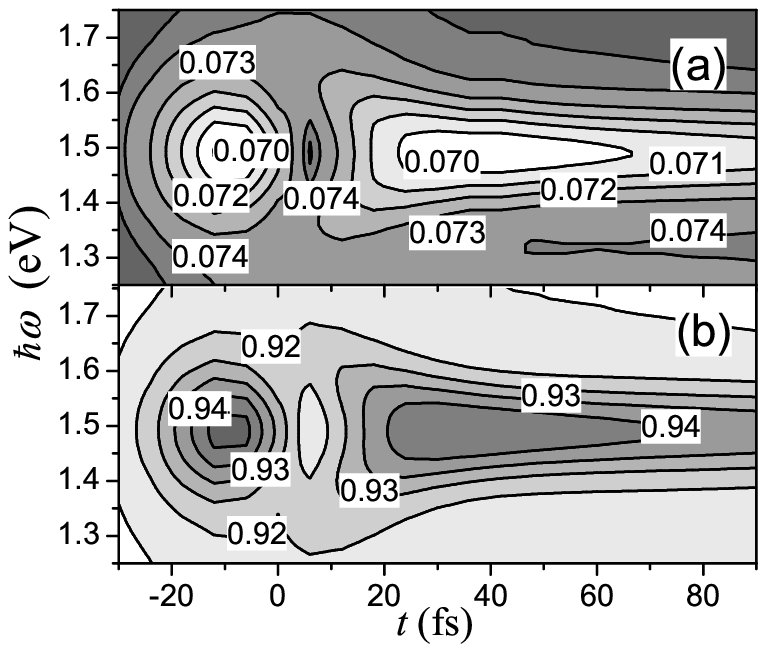}\vspace{-0.7 cm}
\includegraphics{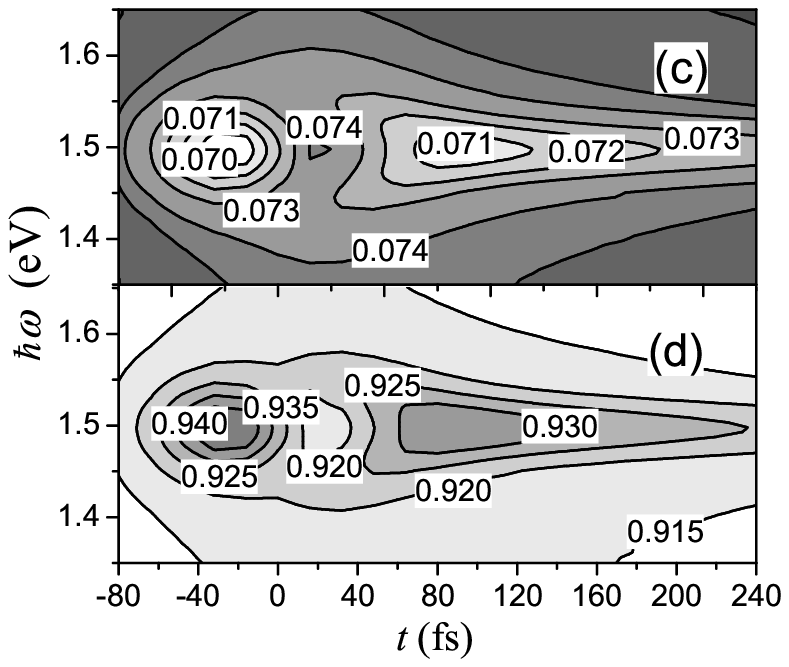}
\end{center}
\caption{Contour plots of reflection (a, c) and transmission (b, d) coefficients
versus energy of probe radiation $\hbar\omega$ and time for pulse durations
$\tau_p=$30 fs (a, b) and 80 fs (c, d) at pumping level corresponding to
${\cal A}_{ex}=3\pi$.}
\end{figure}

\begin{figure}[ht]
\begin{center}
\includegraphics{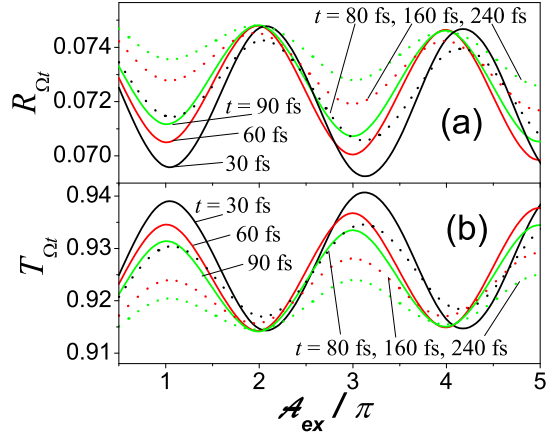}
\end{center}
\addvspace{-1 cm}
\caption{(Color online) Reflection (a) and transmission (b) versus intensity
for $\tau_p=30$ fs and 80 fs (solid and dotted curves) at different delay times
$t=\tau_p$, $2\tau_p$, and $3\tau_p$ (marked) and for $\omega =\Omega$.}
\end{figure}

The reflection and transmission coefficients, $R_{\omega t}$ and $T_{\omega t}$,
are written through the dynamic conductivity as follows: \cite{16}
\begin{eqnarray}
R_{\omega t}=\left|\frac{1-\sqrt\epsilon -4\pi\sigma_{\omega t}/c}
{1+\sqrt\epsilon +4\pi\sigma_{\omega t}/c}\right|^2 ,  \\
T_{\omega t}=\frac{4\sqrt \epsilon}{\left| 1+\sqrt\epsilon +4\pi
\sigma_{\omega t}/c\right|^2} . \nonumber
\end{eqnarray}
Here $\epsilon$ is the dielectric permittivity of a thick substrate and the
geometry of normal propagation of radiation was considered.
The contour plots of $R_{\omega t}$ and $T_{\omega t}$ versus $t$ and $\omega$
are shown in Fig. 5 at pumping level corresponding to ${\cal A}_{ex}=3\pi$
for $\tau_p=30$ fs and 80 fs. Temporal oscillations of response correlate with
the evolution of $\sigma_{\omega t}$ (Fig. 3) due to oscillations of
distribution shown in Fig. 2. Spectral width of Rabi oscillations appears to
be broader for $\tau_p=30$ fs (c.f. Figs. 5a, 5b and 5c, 5d with Figs. 2a
and 2b). The oscillatory behaviors of the reflection and transmission coefficients
at the pumping frequency ($\omega =\Omega$), $R_{\Omega t}$ and $T_{\Omega t}$,
versus pumping level for the delay times $\tau_p$, $2\tau_p$, and $3\tau_p$
are shown in Figs. 6a and 6b, respectively. One can see a few-percent modulation
of response at $t\geq\tau_p$ versus pumping intensity. A visible damping of
this modulation takes place due to the cascade emission of optical phonons,
at $t\geq\nu_{p_\Omega}^{-1}$.

Our calculations are based on the following assumptions. Since the Coulomb
renormalization of interband transitions is weak,  \cite{17} we have used a
single-particle approach. All homogeneous dephasing mechanisms, including
optical phonon emission and carrier-carrier scattering, have been described
phenomenologically, through the dephasing time $\tau_d$ in Eq.(2). An inhomogeneous
broadening due to long-scale disorder is not taken into account, so that
the results are valid for a high-quality graphene. The model collision integral
(3), which is written through an effective phonon energy $\hbar\omega_0$ and
a relaxation frequency $\nu_p$, is used because the only first step of cascade
emission of phonons is essential during the Rabi oscillation process. A detailed
calculations can improve a precision of $v_0$ used in Eq. (3) and will give a
widened peak of distribution at $p_\Omega -p_0$. The above-listed simplifications
of the damping processes do not change the temporal dynamics under consideration
and we have demonstrated that the Rabi oscillations are observable in a typical
graphene sample. The rest of assumptions [the geometry of normal propagation of
radiation, the Gaussian shape of excitation, and the collisionless approximation
used in Eqs. (6)] are rather standard.

Summarizing, we have described the mechanisms of coherent nonlinear response
of an intrinsic graphene under ultrafast interband excitation. The results obtained
demonstrate that the Rabi oscillations, both versus time and versus pumping intensity,
can be easily observed for femtosecond time scales (up to 0.1 - 0.2 ps) at pumping
intensities $\sim$3 - 30 GW/cm$^2$ (which are correspondent to pulse energies
$\sim$10 - 100 nJ for the sport area $\sim 10^{-4}$ cm$^2$).


\begin{thebibliography}{52}
\bibitem{14}
The form-factor $w_t$ is normalized as $\int_{-\infty}^{\infty}dtw_t^2=\tau_p$.
The shape of $w_t$ has little effect on the photoexcitation process because the
response is mainly dependent on $\tau_p$.
\end{thebibliography}
\end{document}